\magnification=\magstep1
\hsize 32 pc
\vsize 42 pc
\baselineskip = 24 true pt
\centerline {\bf{ Subatomic Physics - 100 Not Out and Still Going Strong!}}
\vskip .5 true cm
\centerline {\bf Avinash Khare$^*$}
\centerline { Institute of Physics,}
\centerline { Sachivalaya Marg, Bhubaneswar-751005}
\vfill
* E-mail : khare@iop.ren.nic.in
On the occasion of the birth centenary of the discovery of electron, 
I discuss its role in the area of Elementary Particle Physics.
I emphasize that the discovery of electron marks the end of the speculation
era lasting more than 2500 years. The key developments leading to the discovery
of electron by J.J. Thomson are briefly mentioned. The standard model
is briefly mentioned and it is emphasized that no further progress beyond
standard model is possible unless there is a dramatic advance in 
instrumentation. 
\eject
The year 1897 is truly a landmark year in science since the electron was 
discovered in that year. In the last one hundred years it has
triggered a host of innovations in both science and technology. Volumes
could be written down about the practical applications due to the
understanding of electricity as a stream of subatomic particles
rather than as a continuous fluid. However, to my mind, the greatest
impact of its discovery is in our understanding of the basic
constituents of nature. The discovery of electron marks the end of 2500
 years of speculation about the structure of matter and the beginning
of its current understanding. In this article I wish to offer my perspective
to this momentous advance.

{\bf The Speculation Era}

The question of the basic constituents of nature has attracted
human civilization for a very long time. In our own Indian civilization, 
fire, air, water, sky and earth were
regarded as the five basic constituents. A similar view was also advocated by
Empedocles (490-430 BC) who regarded fire, air, water and earth as the four 
unchangeable elements. The term atom first appears 
in the writing of Democritus in the late fifth century BC. 
According to him, atoms are the smallest parts of matter, though not
necessarily minute. Till almost the end of the nineteenth century,
the Epicurus (341-270 BC) view, shared by prominent physicists, was that
atoms cannot be divided into smaller parts by physical means even though
they have a structure. I might add here that the chemists, by and large, 
were still debating till the late nineteenth century whether atoms were real 
objects or
only mathematical entities which were useful for coding chemical regularities 
and
laws (it is amusing to note that till about 1970, a similar view about 
quarks was held by most particle physicists)!

{\bf Advances in Instrumentation}

It is worth repeating again and again that physics (also chemistry and biology)
is an experimental science and that historically almost all the discoveries 
were
made because of the advances in instrumentation. For example, the three key 
advances
in instrumentation which played a leading role in the discovery of electron 
were (a)
improved vacuum (b) higher voltages (c) cloud chamber. Improved vacua were 
achieved
in the 1850s when J. Geissler (1815-1879) began developing the tubes now
named after him. Soon he was able to reach and maintain pressure of 0.1 mm of
mercury. Michael Faraday experimented with them in 1858 and at one point
commented ``very beautifully wrought''. Higher voltages, on the other hand, 
were
the result of the work of H. R\"uhmkorff (1803-1874) on an improved version of 
the induction coil. By the end of the century, voltages of
the order of 100,000 volts could be generated by these coils. The third major
advance was the development
of a cloud chamber by C.T.R. Wilson (1869-1959) at Cambridge in 1895 for which 
he was awarded the 1927 Nobel prize in physics.
 
It is worth pointing out that neither Geissler nor R\"uhmkorff ever earned
academic degree. Yet they were recognized for their achievement in their 
lifetime. In 1864, R\"uhmkorff won a 50,000 franc prize established by
the French emperor for the most important discovery in the application of
electricity while Geissler was given a honorary degree by the university 
of Bonn in 1868.

One of the biggest drawbacks of the scientific development in India is the 
utter neglect of instrumentation in this country. No wonder, not many 
scientific discoveries have been made in our country.
 
{\bf Towards the Discovery}

Before we discuss the actual discovery of the electron, it is worthwhile to
first discuss a few developments in the eighteenth and the nineteenth century
which led to this discovery.

In 1737, Dufay had written about two distinct electricities which he called 
{\it vitreous} and {\it resinous} and then had proclaimed the law that the 
like charges
repel, unlike charges attract. With the rapid developments in instrumentation,
several major advances took place in the next one hundred years in the area of
electromagnetism. This, in turn, led to the formulation of phenomenological
laws based on new experiments. In particular, exactly one 
hundred years after Dufay's
work, Michael Faraday (1791-1867) gave his famous law of electrolysis. In
modern language his law can be stated as ``the amount of electricity deposited
at  the anode by a gram mole of monovalent ions is a universal constant called 
Faraday constant, and is given by $F=Ne$''. Here $N$, Avogadro number, 
is the number of molecules per mole and $e$ is a universal unit of charge.

How much is $e$? In 1874, G.J. Stoney (1826-1911) obtained the value of $e$ 
to be $\simeq
3 \times 10^{-11}$ esu by using the relation $F = Ne$. Note that $F$ and $N$ 
were reasonably
well known by then. Considering the fact that the present best value of $e$ is 
$4.80\times 10^{-10}$ esu, Stoney's estimate is not all that bad for a first 
and very early try. 

What does this $e$ signify ? In 1881, Helmholtz (1821-1894), in his Faraday
memorial lecture, advocated that if we accept the hypothesis that the
elementary substances are composed of atoms, we cannot avoid concluding that 
electricity (both positive as well as negative) is also divided into definite
elementary portions which behave like atoms of electricity. In 1891, Stoney
baptized this fundamental unit of charge giving it the name {\it electron}. It
is amusing to note that the term electron was actually coined prior to the experimental
discovery of the quantum of electricity and matter which now goes by that
name (curiously enough, the term quark too was coined before the actual experimental
discovery of the constituents of proton, neutron and other hadrons).

{\bf The Discovery}

Let us now turn to the actual discovery of the electron. It was made 
possible by the
study of the cathode rays. One of the major controversies at that time was
regarding the nature of these rays. Two prevalent views were (a)
cathode rays are waves traveling in a hypothetical invisible fluid called the
ether (b) the cathode rays are negatively charged material particles. It is 
worth recalling that at that time many physicists thought that 
the ether was needed to carry light waves through the empty space. The 
feeling was, may be the cathode rays are similar to the light waves. 
On the other hand, 
the first description of electric current as a stream of discrete electric
charges had appeared as early as in 1840s in the work of Fechner and Weber. 
Experiments were clearly necessary to resolve the issue.

{\bf Was Electron Really Discovered in 1897?}

It is stated in numerous books and articles that J.J. Thomson discovered the 
electron in 1897. In fact the whole world (including Resonance) is celebrating
the first century of electron this year. However, I cannot agree that
electron was discovered in 1897. It is of course true that in that year,
Thomson made a good determination of $e/m$ for cathode rays ($e$ being the charge
and $m$ the mass) which surely was an indespensible step towards the identification
of the electron. But then in the same year W. Kaufmann (1871-1947) also made
the same measurement and obtained a good value for $e/m$. Further,
based on his observations, he correctly noted that certain properties of cathode rays are
independent of the nature of the gas they traverse, a clear indication of the
universality of the constitution of the cathode rays. So why is that only
Thomson is given the credit for discovering the electron in 1897 ? Is it 
because he correctly conjectured that the large value of $e/m$ that he had 
obtained (compared
to that of charged hydrogen atom) indicated the existence of a new particle 
(which he called {\it corpuscle}) with a very small mass
on the atomic scale? However, he was not the first one to make that guess ! 
E. Wiechert (1861-1928), even though he was not able to measure $e/m$ for the  
cathode rays, was able to obtain very accurate lower and upper bounds on its
value. On Jan. 7, 1897 he stated his conclusions in a lecture where he 
said ``my experiments on cathode rays show that we are not dealing with the atoms
known from chemistry, because the mass of the moving particle turns out to be
2000-4000 times smaller than that of the hydrogen atom, the lightest of the
known chemical atom''. Historically, this is the first time ever, that a
subatomic particle is mentioned in print and sensible bounds on its mass are
given. However, like Thomson, his conclusions depended crucially in his assumption
about the charge.

So why is J.J. Thomson credited with the discovery of electron ? The point is,
the largeness of $e/m$ for cathode rays (compared to that of charged 
hydrogen atom) did not uniquely settle the issue since it could be from 
largeness of $e$ or smallness of $m$. The issue was conclusively
settled by Thomson in 1899 (and not 1897) when he experimentally measured the
value of $e$ by using the cloud chamber technique developed
only four years before by his student C.T.R. Wilson. Thomson obtained
$e\simeq 6.8\times 10^{-10}$ esu, a very respectable value in view of the
novelty of the method. Using both the results, he quoted a mass of 
$3\times 10^{-26}$ gm 
for the electron, the right order of magnitude. 
Thomson received the 1906 Nobel prize in physics for his 
discovery. 

It may be noted that the electron itself has not exactly turned out
as thought by many people namely as a particle. With the advent of 
quantum mechanics, it is wrong to think that the electron must be 
either a particle or a wave.
The wave-particle duality shows that under certain conditions electron
can act like a particle while under other conditions can act like wave. It 
is one of the irony of the 20'th century physics that the wave nature of 
the electron was in fact 
first shown among others by J.J. Thomson's own son, G.P. Thomson (1892-1975) 
who as a result shared the 1937 Nobel prize in physics.

{\bf The Properties of Electron}

Over the last one hundred years, we have come to know quite a bit about the 
electron. For example, unlike the proton and neutron, it does not experience any 
nuclear force (i.e. strong interaction). 
Further, it has spin ${h\over 4\pi}$ ($h$ being Planck constant),
mass = $9.1\times 10^{-28}$ gm ($\simeq 0.51 MeV/c^2$), negative electric
charge of one unit $(\simeq 4.80 \times 10^{-10}$ esu), non-zero magnetic
moment and lifetime $\tau > 2.7 \times 10^{23}$ yr. There is a good reason 
(local gauge invariance) to believe that {\it a la} energy and momentum, 
the electric charge must also be a conserved quantity in any reaction or a 
decay. As a result, one believes that a free electron must live for ever. 
In case  it turns out that it only lives for a finite life time, that will be 
one of the biggest shock of the twenty first century physics. 

{\bf Birth of Subatomic Physics}

The significance of the discovery of electron can hardly be overemphasized.
Before its discovery, even though many believed in the reality of atoms,
most (if not all) of them shared the view that the atom cannot be decomposed
further. That is why, the discovery of the electron can be regarded as the 
end of
an era which lasted for almost 2500 years and the beginning of the modern era 
of subatomic physics which is continuing for the last one hundred years. 

{\bf Basic Constituents of Nature - The Present Status}

Soon after the discovery of the electron, physicists including Thomson started
building the models of the atom. The most celebrated among them is the Rutherford 
(1871-1933) model of the atom in which the atom consisted of two types of elementary
particles: a light, negatively charged electron and a heavy, positively charged
proton. Only after 1931, when the neutron was
discovered by Chadwick (1891-1974), people realized that there are three 
elementary particles. Soon also came the realization that
there are four basic forces in nature which in the order of decreasing
strength are (i) strong interaction (ii) electro-magnetic interaction
(iii) weak interaction and (iv) gravitational interaction.
Hundreds of the so called ``elementary particles'' were discovered subsequently
mainly due to the use of particle accelerators and bubble chambers. This
created a crisis because the human civilization has always been
fascinated by the idea that the number of basic constituents of nature
should not be too many. The hypothetical quark model was then 
proposed according to which all the hadrons (i.e. mesons and baryons 
which are particles experiencing strong interaction) are made out of still 
smaller constituents
called quarks. Thanks to the beautiful electron-proton inelastic scattering
experiment at Stanford in 1969 (which is analogous to the classic Rutherford 
$\alpha$-scattering experiment), it was  conclusively proved that indeed 
proton and neutron are composite objects which are made out of quarks.

The modern view about the basic constituents of nature and the
interaction between them is described by the so called ``Standard Model''.
According to this model, the basic constituents are six varieties (or flavors) of
quarks (u,d,s,c,b,t) each coming in three colors and six leptons
$(e,\mu,\tau,\nu_e,\nu_{\mu},\nu_{\tau}$), plus their anti-particles. 
Besides, there are twelve gauge bosons, including the photon, eight gluons
and $W^+,W^-,Z^0$ particles. All the quarks and leptons are point objects
with leptons including electron experiencing no strong interaction. On the other
hand, the quarks and gluons, being colored particles, are permanently 
jailed inside the hadrons
and cannot exist as free particles. In this model, the electro-magnetic and 
the weak interactions are unified into 
a single force called the electro-weak force. This is reminiscent
of the unification between the electricity and the magnetism 
by Maxwell (1831-1879). 
One of the crowning glories of the quantum electrodynamics ( which is a 
part of the electro-weak theory) is the spectacular 
agreement between the theory and the experiment about the 
anomalous magnetic moment of the electron to more than seven decimal places. This 
would rank among the highest achievements of the twentieth century physics.

One uncertain aspect of the standard model is the mechanism for generating 
masses
of the quarks, leptons and gauge bosons. A large hadron collider (LHC) is being
built around Geneva in Switzerland to settle this question and one hopes to get 
an answer around the year 2007. An Indian group is also participating in this 
truly international mega project. 

The standard model raises a very disturbing thought. For the first time, in the 
history of mankind, some of the basic constituents of nature do not 
exist as free particles. This raises the issue of the very meaning of the word 
{\it basic constituents}. 

{\bf Beyond the Standard Model}

The standard model, though so successful in explaining all the available
experimental data in subatomic physics, is unable to answer several basic
questions. Further, there is only a partial unification of the basic forces.
In recent years a truly unified theory called the superstring theory has 
been proposed which unifies all the four interactions. One remarkable break 
from the past is that here the basic constituents of nature are not particles 
at all! Rather the basic object is a string of length $10^{-33}$ cm. The quarks,
leptons and gauge bosons are merely the different modes of vibration of the
string. The unification ideas have also brought closer the seemingly contrasting
worlds of the smallest and the largest. In particular, the unification ideas 
hold the promise to explain how the Universe evolved after the big bang.
Another possibility is that the quarks, leptons and other particles of the
standard model are themselves composed of more elementary objects. I must
make it clear here that so far we have no experimental evidence for
any of the ideas beyond the standard model.

{\bf Electron as a Probe in Subatomic Physics}

Over the years, the electron has played a big role in uncovering the mysteries 
of the subatomic physics. This is primarily because it is so light, 
experiences no strong interaction, has non-zero charge and lives for ever. 
Its first major use was in deep inelastic scattering experiments 
where it was used as a probe to uncover the structure of the proton and other 
hadrons. Its major use, however, was in discovering new particles in the 
electron-positron colliders. The big advantage here is that the strong interaction 
background is absent in these collisions. The future progress in subatomic 
physics crucially depends on whether we are able to build a very high energy 
electron-positron linear collider or not.

{\bf Subatomic Physics - a Hundred Year Hence}

Finally, what will scientists think a century later ? What concepts which we
hold dear today will be regarded as the ether of 1997? What will be the
elementary particles of the twenty first century? Would there be an ultimate
theory of everything (TOE) or as others would like to say, a truly unified theory
(TUT) ?

One very disturbing aspect is that the recent theoretical activities are highly
speculative with absolutely no experimental data to back them up. I believe
that it is highly dangerous to rely too much on formal elegance and too 
little on facts.
Unless the scientific community is willing to put higher emphasis on 
instrumentation, 
I am afraid, there will be a big pause in the dialogue between experiment and  
theory. Hopefully, our children and grand children will be wiser than our generation.

Finally will we ever have a TOE or TUT ? I find that on the whole, the western 
civilization would answer this question in the affirmative though there may not 
be any unanimity on whether it would take 20, 100 or one million years. However, I 
completely {\it disagree} with this view point. 
I believe that quarks and leptons
represent merely yet another Sari of Draupadi.

{\bf Suggested Reading}

\item {.} A Pais, {\it Inward Bound}, Oxford Univ. Press (1986).

\item {.} 100 years of Elementary Particles, Beam Line {\bf 27}, No.1
(Spring 1997), Stanford Linear Accelerator Center.

\item {.} For a memorial guided tour consult the website; http://www.aip.org/
history/electron

\item {.} For a semi-popular modern view of the subatomic physics, see for 
example, J. Bernstein, {\it The Tenth Dimension}, McGraw-Hill Publishing Company 
(New York, 1989). 

\vfill
\eject
\end